Improving Policy-Oriented Agent-Based Modeling with History Matching: A Case Study

David O'Gara*[1], Cliff C. Kerr[2], Daniel J. Klein[2], Mickaël Binois[3], Roman Garnett[1,4], Ross A. Hammond[1,5,6,7]

**Affiliations**
[1]Division of Computational and Data Sciences, Washington University in St. Louis, St. Louis, MO
[2]Institute for Disease Modeling, Bill & Melinda Gates Foundation, Seattle, WA
[3]Acumes team, Université Côte d'Azur, Inria, Sophia Antipolis, France
[4]Department of Computer Science, McKelvey School of Engineering, Washington University in St. Louis, St. Louis, MO
[5]Public Health, Brown School, Washington University in St. Louis, St. Louis, MO
[6]Center on Social Dynamics and Policy, Brookings Institution, Washington, DC.
[7]The Santa Fe Institute, Santa Fe, NM
*To whom correspondence should be addressed: david.ogara@wustl.edu

**Abstract**

Advances in computing power and data availability have led to growing sophistication in mechanistic mathematical models of social dynamics. Increasingly these models are used to inform real-world policy decision-making, often with significant time sensitivity. One such modeling approach is agent-based modeling, which offers particular strengths for capturing spatial and behavioral realism, and for *in-silico* experiments (varying input parameters and assumptions to explore their downstream impact on key outcomes). To be useful in the real-world, these models must be able to qualitatively or quantitatively capture observed empirical phenomena, forming the starting point for subsequent experimentation. Computational constraints often form a significant hurdle to timely calibration and policy analysis in high-resolution agent-based models. In this paper, we present a technical solution to address this bottleneck, substantially increasing efficiency and thus widening the range of utility for policy models. We illustrate our approach with a case study using a previously published and widely used epidemiological model.

**Introduction**

The mitigation and suppression of infectious disease spread remains one of the grand challenges in our society. For more than two decades, scientists have developed complex simulation models to study, describe, and ultimately intervene on disease dynamics (Burke et al. 2006; Ferguson et al. 2006; Germann et al. 2006; Halloran et al. 2008; Longini Ira M. et al. 2005). During the initial and subsequent waves of the COVID-19 pandemic, a number of sophisticated agent-based ('individual-level') models were developed to forecast infectious disease spread, inform policy planning, and explore counterfactual scenarios (Aylett-Bullock et al. 2021; Ferguson et al. 2020; Hammond et al. 2021; Hinch et al. 2021; Kerr, Mistry, et al. 2021; O'Gara et al. 2023; Ozik et al. 2021). Before use for policy analysis, a key step in developing an epidemiological agent-based model is conducting a series of tests to ensure it can reproduce observed epidemic trends, a process known as calibration. One of the major challenges in model calibration is that the simulation process is often computationally expensive due to the high level of detail encoded in these models, meaning that extensive experimentation is often not feasible under short time horizons. One model that received widespread usage during the pandemic, while facing these calibration challenges, was Covasim (Kerr, Stuart, et al. 2021). Like all scientific models, Covasim makes several necessary simplifying assumptions and relies on imperfect data to simulate disease spread in as realistic manner as possible. Due to Covasim's high visibility and direct impact on policy response planning in both the United States and worldwide, a key goal throughout the project has been to understand the model's ability to reliably reproduce observed epidemic trends, and to analyze the model's sensitivity to aforementioned assumptions and data.

Here, we use the example of Covasim to introduce a novel approach for calibration (analysis of Covasim's ability to re-create simulated disease spread in a structured population). We conduct several rounds of 'history matching,' a Bayesian approach that aims to rule out regions of the model parameter space that are unlikely to give rise to the observed data, followed by calibration via approximate Bayesian computation (ABC). History matching has been applied to a wide variety of complex simulation models, including ones of galaxy formation (Vernon, Goldstein, and Bower 2010), HIV spread (Andrianakis et al. 2015), systems biology (Vernon et al. 2018), and HPV spread (Iskauskas, Cohen, et al. 2024). In particular, history matching has also been done for the JUNE model in England (Vernon et al. 2022), and Covasim has been calibrated using ABC in a different context, but these tools have not been combined for COVID-19 epidemiological models. Since the data generating process is expensive for mechanistic agent-based simulations, we utilize an emulator model (via heteroskedastic Gaussian process regression (O'Gara 2024b)) to explore the parameter space and rule out unpromising potential design locations, and ultimately aid in the calibration.

Our point of departure is the calibration analysis in the first major publication featuring Covasim (Kerr, Mistry, et al. 2021). In that study, the authors calibrated four uncertain model parameters and posed the calibration search as a hyperparameter optimization problem. This approach was able to generate many model runs consistent with observed data, but had two potential drawbacks: (1) the calibration required over 100,000 model runs, corresponding to nearly 35 days of computing time; (2) the hyperparameter optimization approach could make uncertainty quantification somewhat challenging, as has been discussed in other work (Panovska-Griffiths et al. 2023).

Our work seeks to supplement the analysis by Kerr et. al by identifying model parameterizations that are most likely to match empirical data (in this case, the time-series of diagnoses and deaths) across random seeds. In the original Covasim publication, the goal of the calibration problem was to acquire a large number of model runs that satisfied empirical data, building up a sort of "library" of potential parameter and random seed pairings for which to explore policy interventions. Other authors employing related methodologies have referred to this type of approach as "trajectory-oriented optimization" (Fadikar et al. 2023). This approach is valuable in that it does not make assumptions about the mean behavior of repeated model runs, and potentially offers a wide range of parameters that might satisfy the data for a set of random seeds, making claims about distributions of eligible model parameterizations is challenging, as well as informing novel geospatial contexts, such as another city. Further, to precisely reproduce a stochastic model's results, one would also have to save the runtime environments and random number streams. While this is possible with standard computing tools, such as version control and containerization, in high-visibility epidemiological modeling that can (and in the case of Covasim, did) inform disease response planning, it is important to make a distinction between the mathematical model, the software implementation that compiles its results, and the fact any model is an approximation of the real world, which is itself one stochastic realization of an underlying process. This means that some stochasticity is fundamentally irreducible (also known as "aleatoric uncertainty"). Thus, identifying model parameters that are expected to yield outputs comparable with observed data, regardless of the random number stream, is a useful goal. Perhaps just as important is the variance of the outputs induced by model parameters that satisfy empirical data, providing a notion of the stability of model results due purely to stochasticity, and even "good" model parameters may produce realizations incompatible with data. We also note that these approaches are not mutually exclusive, since an initial set of experiments could be conducted to set initial bounds on model behavior, before running a more detailed calibration.

Our model calibration goals are as follows: (1) Many models focus on one 'best' parameterization. We are interested in ruling out large regions of the uncertain model parameter space, and from this reduced space, estimating a posterior distribution of acceptable parameterizations; (2) for this model posterior, we are interested in the uncertainty quantification of the acceptable set of parameters and (3); we want to limit the number of total model repetitions, as the data-generating process is expensive for complex stochastic epidemiological models.

Our results demonstrate that high-fidelity emulation of Covasim is possible via history matching. Our four rounds of history matching ruled out over 99% of the parameter space volume, from 40 points per dimension for four parameters (2,560,000 points total) to 21,114 points. This allowed us to then conduct an approximate Bayesian computation (ABC) analysis with more informative priors which allowed for fine-grained exploration of a much smaller region of parameter space. The resulting accepted posterior distributions are able to match empirical data across random seeds. We were also able to limit the number of total simulations, by using 5,300 model runs, an improvement over previous analyses. Our work has implications for pandemic preparedness analysis, and is an important comparator to other calibration paradigms, such as the trajectory-oriented optimization approach. In high-stakes epidemiological modeling aimed to inform disease containment, we argue that multiple model alignment strategies should be explored, which will hopefully better inform potential policy response.

## Methods

*Covasim*

Our setting is the Covasim calibration problem in Kerr et. al 2021. Covasim's functionality is described in detail in both Kerr et. al 2021 and the methods paper (Kerr, Stuart, et al. 2021), but we briefly describe the core components here. Covasim is a large-scale agent-based stochastic transmission model. The present analysis with Covasim version 2.1.2 uses a model population of 225,000 agents with dynamic rescaling to represent the 2.25 million residents of King County (i.e. the Seattle metro and its surrounding area). The model also uses empirically informed contact networks from the Synthpops (Mistry and Kerr 2021) library, with four contact layers representing home, workplace, school, and community contacts. Like other agent-based COVID-

19 models, Covasim follows the Susceptible-Exposed-Infected-Recovered-Dead (SEIRD) model, with their discrete state describing their current health status, which also determines the probability of when and which state they will transition to. The model also contains detailed within-host dynamics modeling an agent's disease progression (for example, allowing agents transmission probability to peak the day before or day of symptom onset). The key step of the model is computing the new infections from infectious agents to susceptible ones. An agent's probability of infection may be modulated by one or more factors depending on the context and that of their network contacts (for example, whether a stay-at-home order is in place, lowering the probability of transmission between community contacts).

During the calibration in the original publication, the authors varied four uncertain model parameters: the per-contact transmission rate per day "*beta*", the relative per-contact transmission reductions in work and community layers from March 23, 2020 onwards ("*bc_wc1*"), the relative per-contact transmission reduction in long-term care facilities from March 23, 2020 onwards ("*bc_lf*") and the odds ratio of symptomatic agents taking a diagnostic test ("*tn*"). Each of these parameters shape the overall model behaviors and dynamics, and thus, the authors conducted a set of calibration runs to arrive at a representation of Covasim's baseline dynamics for the Seattle metro area. As discussed in the original work, each of these model parameters were given a search range which the authors used as uniform priors and posed the model calibration process as a hyperparameter optimization problem, minimizing a weighted scalar loss function (the "mismatch") between model outputs (for example, the time-series of diagnoses and deaths and the age distribution of infections) and observed data. The original work used Optuna (Akiba et al. 2019), a publicly available hyperparameter optimization library, with the structured Tree Parzen Estimator (TPE) sampler (Bergstra et al. 2011). The original calibration used an acceptability threshold of 30 for the mismatch function, finding 15,092 simulations satisfying this cutoff (and another 8,821 without using empirically informed mobility data).

In addition to trajectory selection, it is valuable to estimate how Covasim will perform at novel proposed input settings on new random seeds. Repeated model runs across random seeds are critical to understanding the behavior of a stochastic system, as replication is a canonical approach to reliably separate signal from noise (Baker et al. 2020; Gramacy 2020; Santner, Williams, and Notz 2018). Uncertainty quantification (UQ) of stochastic systems can be challenging without repeated model runs at the same input locations. Understanding the average behavior of a stochastic system is especially important when the system under study is being used to inform the efficacy of potential policy interventions, as in the case of Covasim, which has been used in over a dozen countries (Kerr, Stuart, et al. 2021).

*Emulation and History Matching*

Like many detailed computer simulation models, the data-generating process for Covasim is somewhat expensive, requiring about 30 seconds per model run. Thus, extensive large-scale simulation experiments are not possible in a reasonable amount of time, which motivates the use of an emulator (or surrogate) model, a computational approximation of the true underlying computer model, but can be evaluated more efficiently, often by several orders of magnitude (Gramacy 2020). History matching is a technique that uses emulator models to rule out regions of the parameter space which are unlikely to match empirical data (Andrianakis et al. 2015; Iskauskas, Vernon, et al. 2024; Vernon et al. 2022, 2010).

More concretely, suppose we have observed data $Y$ corresponding to stochastic simulator output $g(\theta)$ for a vector of model parameters $\theta$. We assume the data generating process:

$$Y = g(\theta) + \sigma_{MD}^2(\theta) + \sigma_\epsilon^2,$$

where $\sigma_{MD}^2(\theta)$ represents the model discrepancy (accounting for any systematic deviations between simulation and reality) and $\sigma_\epsilon^2$ represents observational noise. We introduce an emulator function $\hat{g}(\theta)$ to estimate the behavior of $g(\theta)$. After running the model $N$ times, we can estimate the mean $\hat{\mu}_g(\theta)$ and variance $\hat{\sigma}_g^2(\theta)$ of the model output at unseen parameterizations. Using the observed data, our estimates, and uncertainties, we rule out a potential model parameterization via an *implausibility measure* $I(\theta)$ and remove them if they are above the cutoff criteria:

$$\frac{|Y - \hat{\mu}_g(\theta)|}{\sqrt{\hat{\sigma}_g^2(\theta) + \hat{\sigma}_{MD}^2 + \hat{\sigma}_\epsilon^2}} \geq I(\theta)$$

Specifically, since the system at hand contains multiple outputs, we use the maximum implausibility measure across Q outputs $I_M(\theta) = \max_{i \in Q} I_i(\theta)$. Any remaining points after applying the cutoff are deemed "Not Yet Ruled Out" (NROY) (Iskauskas, Vernon, et al. 2024). Given that many trajectories shown in Kerr et. al 2021 were able to satisfy empirical data, we assume no model discrepancy. However, estimates of model discrepancy could also be elicited via expert input (Vernon et al. 2010), by developing a separate model of the discrepancy (Holthuijzen et al. 2024), or as an additional multiplier on the overall uncertainty in the model (Andrianakis et al. 2015). In the case of our history matching approach, this also corresponds to removing more potential input locations, since adding a term to the denominator would deflate the implausibility measure (Andrianakis et al. 2015). Model discrepancy could also be revisited under the condition where NROY space is estimated to be empty (stated differently, if we were not able to identify non-implausible model parameters, this would require revisiting both the model itself and the model discrepancy) (Vernon et al. 2022). We also assume no observational error since most data sources in Covasim were from publicly available sources, such as dashboards, and could be updated in near real-time and we apply weekly smoothing to time-series input data. Further, since our model outputs are modeling the observational process (based on reported diagnoses and deaths), we assume any under-reporting would get absorbed into the symptomatic testing odds ratio parameter. The implausibility measure essentially rules out parameterizations that simulate outputs not sufficiently close to the data of interest after accounting for model (and observational) uncertainty. A common cutoff value $I(\theta)$ is 3, inspired from (Pukelsheim 1994) showing that at least 95% of any continuous unimodal distribution is contained within three standard deviations and has been used in the history matching literature (Andrianakis et al. 2015; Iskauskas, Vernon, et al. 2024; Vernon et al. 2022, 2010).

*Emulation via hetGPy*

Our chosen emulation strategy is heteroskedastic Gaussian process (hetGP) modeling, following the approach of (Binois, Gramacy, and Ludkovski 2018), and implemented via the Python package hetGPy (O'Gara 2024b). The choice of a hetGP-style methodology allows us to relax the assumption of a stationary stochastic process: namely, that we allow the noise to vary throughout the input space. Since we are modeling the relationships both across and within parameters (for example, the number of diagnoses could be quite noisy if the transmission rate is low and the symptomatic testing odds ratio is high), this assumption is well-grounded. The hetGP approach has been used in other epidemiological modeling contexts (Fadikar et al. 2023; Ozik et al. 2021; Reiker et al. 2021; Shattock et al. 2022), but has not yet been used to analyze Covasim. The practical motivations also complement the methodological: in the hetGP paradigm, where we have a dataset of $N$ observations that are comprised of $n$ unique parameterizations with $\{a_1, \ldots a_n\}$ repeated runs at each parameterizations such that $\sum_{i=1}^{n} a_i = N$, emulator training and inference is on the order of the unique design locations $O(n^3)$ rather than the full dataset $O(N^3)$ which is required in the standard Gaussian process (GP) framework (Garnett 2023; Gramacy 2020). This is accomplished by several applications of matrix identities (known as the Woodbury identity (Binois et al. 2018; Gramacy 2020; Harville 1997)) which allow matrix determinants and inverses to be calculated as a function of $O(n^3)$, which are the most crucial steps in Gaussian process maximum likelihood estimation. In our context, model replicates are essential to separate signal from noise, and thus we are able to perform 20 replicates or more at each design location without having to incur the "full-$N$" emulator training cost these model runs. The choice of replicates was determined as a function of wanting at least at least 10 parameter locations per dimension (in this study, four) and then as a function of available computing resources. Earlier versions of this analysis also explored using more parameterizations with fewer replicates, but we found that 20 replicates modeled the variance more robustly, and is similar to approaches employed in other work (Andrianakis et al. 2015).

*Sequential Design*

We begin our first wave of history matching with a maximin Latin hypercube design of 50 model parameterizations, with 25 replicates at each design location, and at subsequent waves, used 20 replicates per design location. In the initial round, we construct emulators for two outputs cumulative diagnoses and cumulative deaths at three model timesteps representing April 2, April 26, and June 8, 2020. We also emulate the number of active infections on March 26 and May 7. Taken together, these outputs estimate the total size of the epidemic, as well as a snapshot of the peak of infectious agents. We then estimate the mean and variance across a dense grid of 40 input locations for each of the four model parameters, corresponding to a total of $40^4 = 2,560,000$ candidate model parameterizations, which would correspond to nearly 2.5 years of computing time using the simulator, but can be evaluated via emulation in a matter of minutes. Following the removal of implausible inputs at each round, we then draw a sample of 50 new designs with a maximin strategy (Sun and Gramacy 2024) over current NROY space and simulate 20 replicates each using Covasim. As rounds continue, we train more accurate emulators (by emulating a higher density of time steps) over a smaller region of parameter space, and thus also include stricter implausibility thresholds to remain in NROY space. The history matching rounds are summarized in **Table 1**.

| Wave | Model Outputs (timesteps/dates) | Cutoff $I(\theta)$ | NROY Samples | % of Parameter volume |
|---|---|---|---|---|
| 1 | *Cumulative Diagnoses:* Apr-02, Apr-26, Jun-08<br>*Cumulative Deaths:* Apr-02, Apr-26, Jun-08<br>*Active Infections:* Mar-26, May-07 | 3.0 | 184,974 | 7.23% |
| 2 | *Cumulative Diagnoses:* Apr-02, Apr-26, Jun-08<br>*Cumulative Deaths:* Apr-02, Apr-26, Jun-08<br>*Active Infections:* Mar-26, Apr-19, May-07<br>*Diagnoses:* Apr-02 | 3.0 | 142,813 | 5.58% |
| 3 | *Cumulative Diagnoses:* Apr-02, Apr-26, Jun-08<br>*Cumulative Deaths:* Apr-02, Apr-26, Jun-08<br>*Active Infections:* Mar-26, Apr-19, May-07<br>*Diagnoses:* Apr-02 | 2.7 | 54,848 | 2.14% |
| 4 | *Cumulative Diagnoses:* Apr-02, Apr-26, Jun-08<br>*Cumulative Deaths:* Apr-02, Apr-26, Jun-08<br>*Active Infections:* Mar-26, Apr-19, May-07<br>*Diagnoses:* Apr-02, Apr-26 | 2.5 | 21,114 | 0.82% |

**Table 1:** Summary of history matching rounds.

*Detailed calibration via hetGPy and Approximate Bayesian Computation*

Upon ruling out large region of the model parameter space, we expect that the remaining parameters may contain good fits to empirical data on average. On this smaller space, we train two final hetGPy emulator models on the time-series of new diagnoses and deaths, respectively, and calibrate to empirical data using approximate Bayesian computation (ABC). ABC is a likelihood-free inference method which can be used when the likelihood of a data-generating process is intractable but can be simulated from (as in the case of stochastic agent-based models). The core idea of ABC is that we can keep parameters that generate outputs similar to observed empirical data, calculated via a summary statistic or distance metric. With a sufficient number of simulations, our posterior samples will then approximate the observed data (Baker et al. 2020; Beaumont et al. 2009; Sisson, Fan, and Tanaka 2007).

In our ABC analysis, we define informative truncated normal priors over the remaining NROY space based on the mean and standard deviation of the remaining NROY samples after four rounds of history matching, since the remaining NROY samples could be at least reasonably approximated by their mean and standard deviation. We then draw sample simulations from our hetGPy via Sequential Markov Chain Monte Carlo, using 5 chains with 2,000 samples each for 10,000 posterior samples total. We conducted the ABC analysis with the Python package PyMC (Abril-Pla et al. 2023), utilizing an epsilon value of 5, an independent Metropolis-Hastings proposal, and a Gaussian distance metric.

*Software Availability*

All analyses in this work used Covasim version 2.1.2, using Python 3.10. Simulations were run on the Washington University RIS compute platform. All analysis scripts are available at https://github.com/davidogara/covasim-calibration/, and the Docker image to reconstruct our analysis is available at https://hub.docker.com/r/dogara/covasim-py310. Data are also archived via Zenodo at 10.5281/zenodo.14574663 (O'Gara 2024a).

## **Results**

*History Matching Rules Out Large Regions of Parameter Space*

The progression of simulator runs across history matching rounds in shown in **Figure 1**. We see model runs become more accurate across subsequent rounds, especially for the time-series of diagnoses. We also observe lower variance in model outputs in later rounds. Fitting the time-series of deaths is challenging, due to the low number of deaths per day reflected in empirical data, but our fits to the cumulative number of deaths improves with increased rounds.

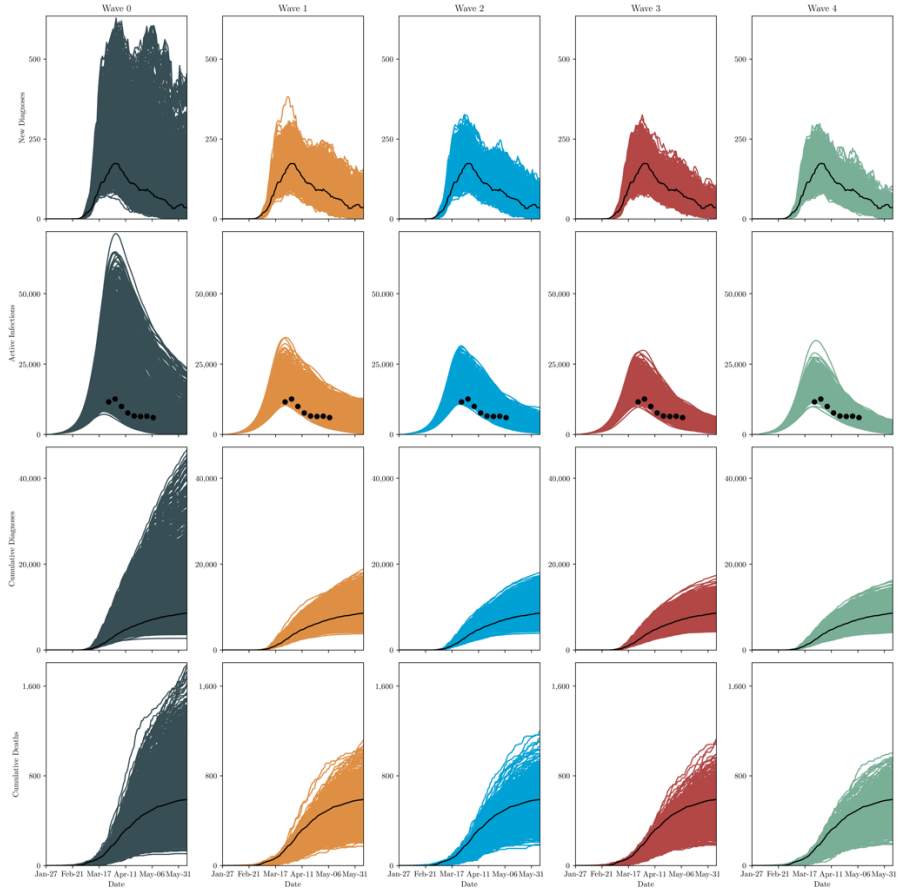

**Figure 1:** Summary of Covasim model runs across rounds of history matching. Each color denotes a history matching round, where gray corresponds to the first, and green corresponds to the fourth.

Examination of the non-implausible space via pairwise plots reveals the reduction in parameter space volume. **Figure 2** shows the optical depth plots for our parameter pairs: for each point in the parameter grid, we plot the proportion of the remaining NROY volume that satisfies each parameter pair, showing where there may still be large regions of potential model parameterizations that could satisfy the data. The two most apparent takeaways are the reduction for the *tn* parameter (modeling the odds ratio of symptomatic agents to take diagnostic tests compared to non-symptomatic), and the clear inverse relationship between the overall transmission rate *beta*, and the transmission reduction in work and community contacts (*bc_wc1*). This is expected that as the transmission rate increases (increasing the overall probability of infection), the transmission reduction for work and community contacts must also decrease to calibrate to empirical data.

We do not see the same type of structure in the comparison between the transmission rate *beta* and transmission reduction in long-term care facilities *bc_lf*, but this is not surprising given that infections in long-term care facilities only accounted for 1.2% and 0.1% of overall transmissions before school closures and after the stay-at-home policy in the analysis by Kerr et. al 2021, compared to 58.6% and 52.0% in work and community contexts during the same period.

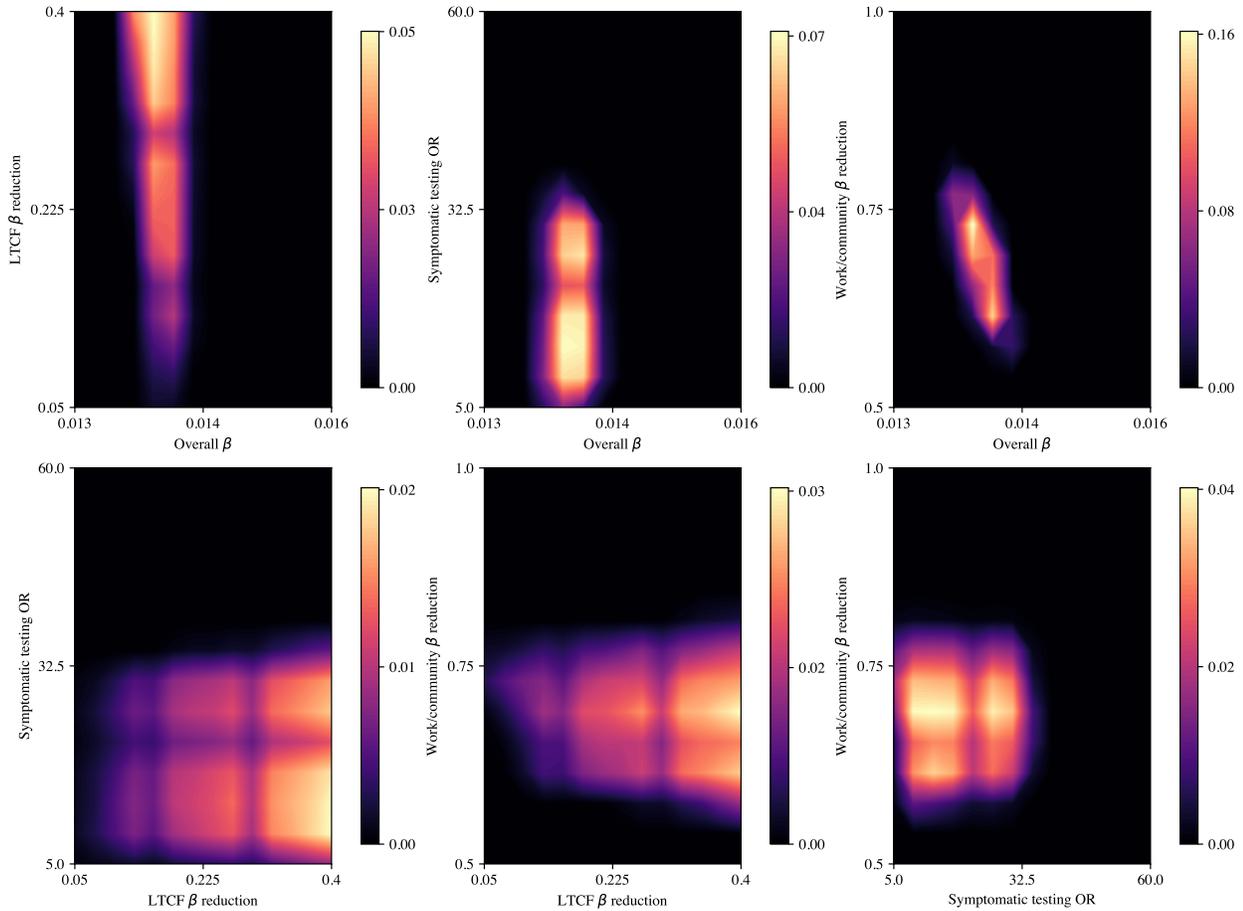

**Figure 2:** Optical depth plots of parameter pairs after three waves of history matching. Plots are generated by calculating the proportion of non-implausible points in parameter space that share each x,y coordinate, estimating the remaining the remaining volume of NROY space.

As history matching rounds continue, we are able to train more accurate emulator models because we are searching over a much smaller region of parameter space. At each round, we choose new design locations to span NROY space while also adhering to a maximin strategy (Sun and Gramacy 2024). **Figure 3** shows the progression of design locations (50 per round, 20-25 replicates at each design location), and we see that we are able to concentrate our chosen designs to span smaller and smaller regions of interest.

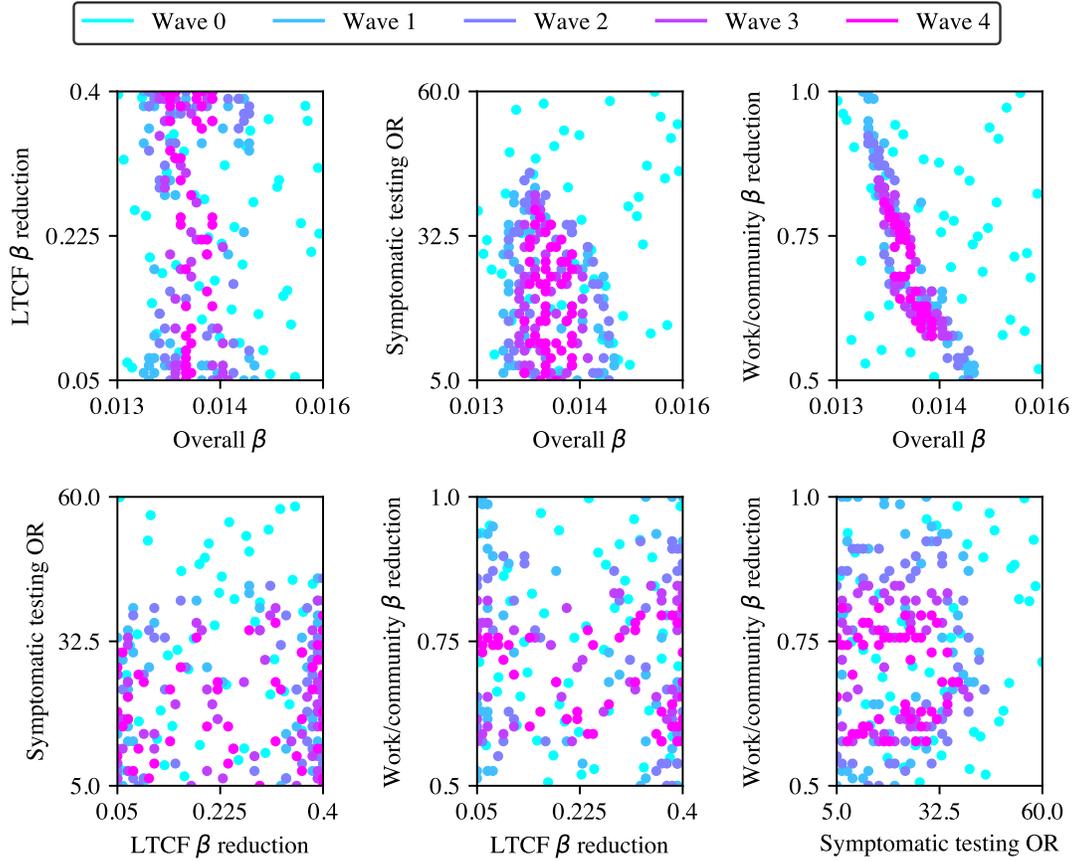

**Figure 3:** Progression of design locations (Covasim parameter simulations) across history matching waves. Note that "Wave 4" is the final round of history matching, which were the design locations for the ABC analysis.

*High-Fidelity Out of Sample Matching is Possible with ABC Following a History Match*

After three rounds of history matching, we removed over 99% of parameter space from consideration. The remaining parameter volume across rounds is shown in **Table 1**. With a substantial reduction of potential model parameterizations, it is then feasible to train detailed time-series emulation models and calibrate them to empirical data using Approximate Bayesian Computation (ABC).

We tested the accuracy of our posterior distribution by simulating 50 samples. Our fits to empirical data are shown in **Figure 4**. We are able to estimate the time-series of diagnoses and death (two of the metrics shown in Kerr et. al 2021) without re-running simulations on the same seeds they were trained on. In this case, we used new random seeds during each history matching wave and for the final 50 posterior simulations. Specifically, each random seed for a simulation corresponds to the order in which simulations were run (in this case, 0 to 5,350). Fitting to the number of estimated active infections is more challenging, especially in the case of limited observations. We also see that our 50 and 90% quantiles are symmetric about the median model run. Thus, we expect stable results on other random seeds, indicating that this model parameterization is largely invariant to stochastic model behaviors from one random seed to another. We also note that our calibrated model runs did not directly rely on an emulation of the number of active infections (due to limited data), but these data were used in the history matching process to prevent the number of diagnoses and projected infections from diverging from one another. The fits to age distributions of cumulative diagnoses and deaths are shown in **Figure S1**. MCMC diagnostics for the ABC-SMC experiment are available in **Figures S2-S3**.

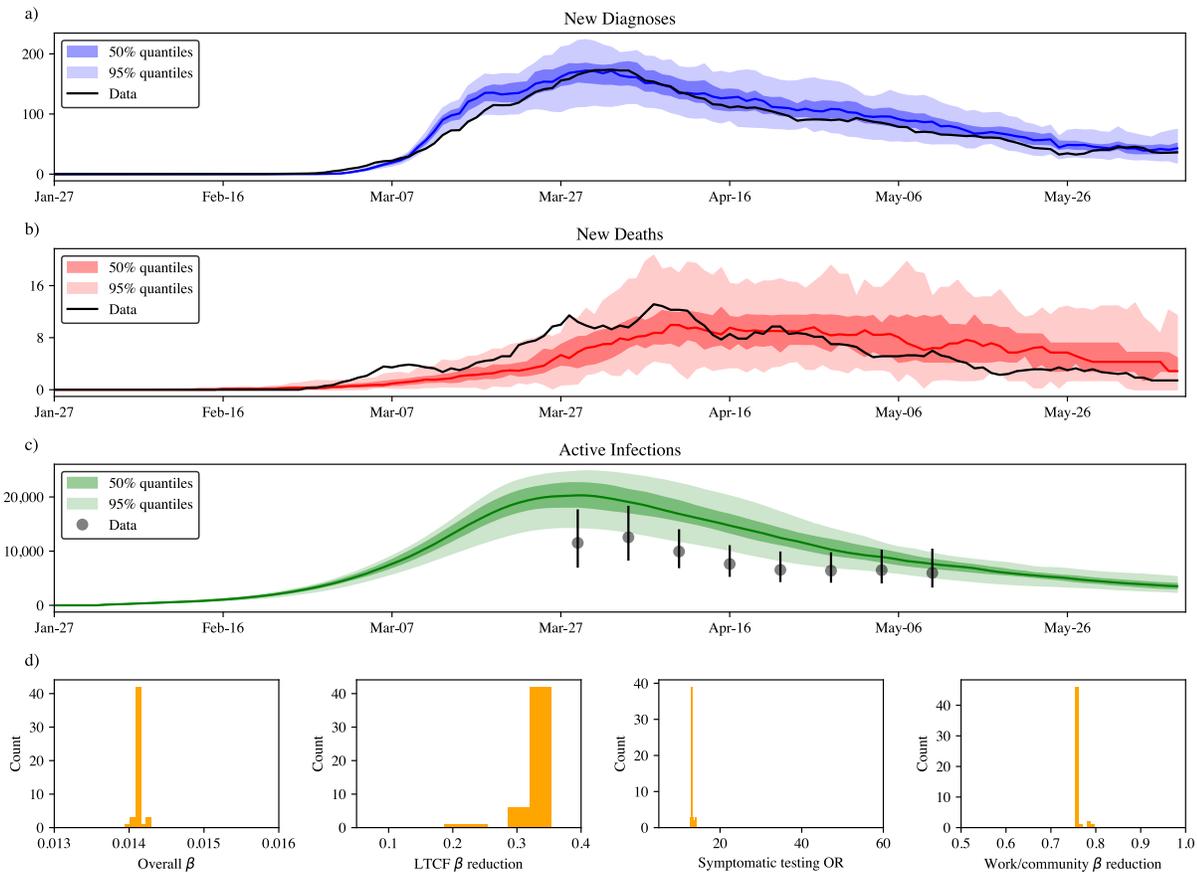

**Figure 4:** Fit to empirical data after history matching and emulation with hetGPy. hetGPy models were fit to the time-series of diagnoses and deaths and calibrated to empirical data via Sequential Monte Carlo Approximate Bayesian Computation. Panel (d) shows the 50 posterior samples that created the results in (a)-(c).

*The Effect of Policy Interventions*

Calibration of an ABM is critical to ensuring that our simulation is a reasonable approximation of the real world (Hammond 2015), especially when it is used to inform disease response planning, because the model can be used to conduct an *in-silico* experiment where we assess the effect of policy interventions on disease spread compared to baseline conditions. In the case of Covasim, the study by Kerr et. al explored the effect of multiple interventions, one of which was a counterfactual scenario in which King County rapidly expanded its testing, contact tracing, and quarantine programs on June 1, 2020, following the lifting of a stay-at-home directive. Crucially, the authors found that a high level of testing and contact tracing would have substantially curtailed disease spread, especially in regard to the estimated number of active infections in the population in summer 2020. Our analysis in **Figure 5** also comports with these findings when we use our calibrated model parameters and project them forward in time. We do observe that the simulations from the original work (labeled TTQ) appear to follow the data more closely, while our method (ABC) contains a sharp spike in infections following the lifting of the stay-at-home order, followed by a more rapid decline in diagnoses. We also note that both of the "status quo" experiments estimate similar levels of active infections in the population from June 1, 2020, onward.

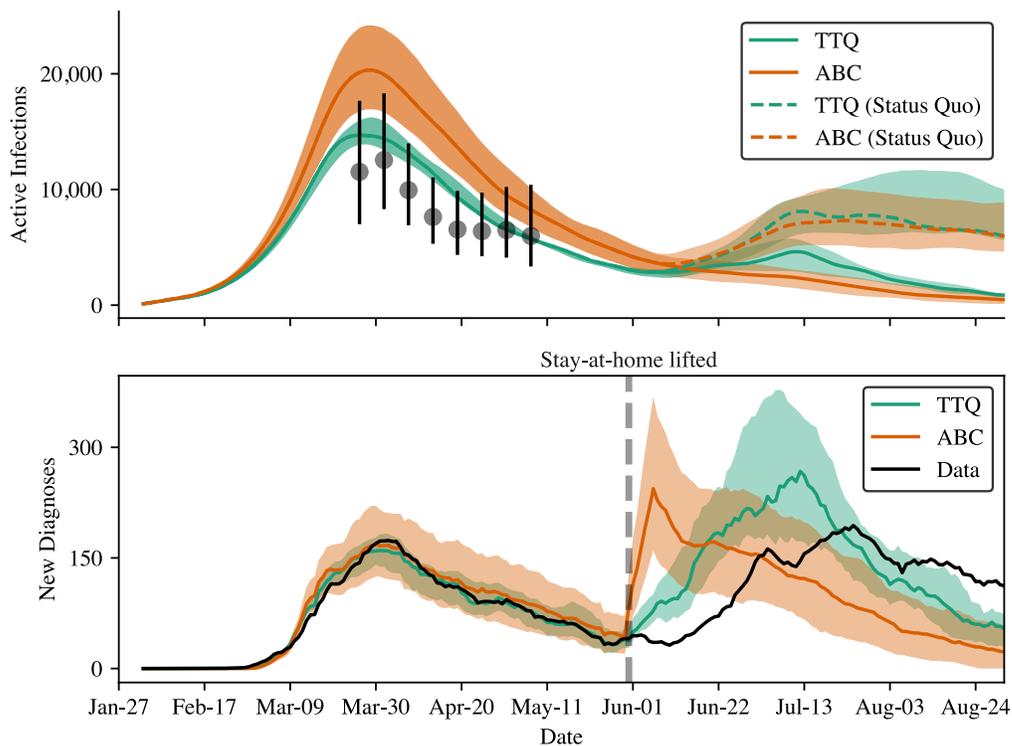

**Figure 5:** Calibrated model parameters are capable of informing policy intervention scenarios. Simulations reproduce a counterfactual policy intervention discussed in Kerr et. al where King County expanded its diagnostic testing and contact tracing capacity to allow re-opening in summer 2020 while avoiding increases in infection and diagnoses in the population. TTQ refers to the original "trace-test-quarantine" policy strategy simulated in Kerr et. al, and ABC refers to the same intervention, except using the calibrated parameters from Approximate Bayesian Computation. The "status quo" simulations refer to the simulations run with the TTQ and ABC parameters, but without policy interventions.

Like all mathematical models, Covasim (and subsequent analyses we conducted with it) are limited by the fact that we are attempting to represent highly complex human behavior in a necessarily simplified (modeled) form. We recognize that by prioritizing fits to some epidemic trends (such as reported cases and deaths) we may sacrifice fits to other metrics (such as the reported number of active infections in the population).

**Discussion**

We see at least three direct extensions of this work. The first is that advancements in model calibration allow revisiting of fixed (non-calibrated) model parameters, allowing modelers to broaden their parameter search space when trying to align models to data. Efficient calibration can serve as a guide to help enrich existing model behaviors: in the case of our Covasim analysis, it is clear that the odds ratio of symptomatic vs non-symptomatic agents taking a diagnostic test has a large effect on reported cases. This mechanism serves as a proxy for health-seeking behavior, and could be extended to accommodate richer behavioral mechanisms, which recent modeling work has called for (Bedson et al. 2021). The space of all model assumptions and potential functional forms of model behaviors is a far larger (perhaps infinite) decision space, which could also have to be explored in calibration. Finally, in the case of Covasim, advances to its calibration could benefit other models in the "Starsim" suite (Kerr et al. n.d.), which share many aspects of the Covasim framework, meaning these methods could be applied to calibrate a broad class of models of health and disease.

More broadly, the work presented here is intended to benefit high-stakes policy-oriented agent-based modeling across many topic domains, helping to "make evidence go further" by increasing the utility of simulations to inform counterfactuals (Hammond and Barkin 2024). To maximize utility to decision-makers and overall trust in such models, analysis of model dynamics, and uncertainties therein, is vital. Above we focused on the specific case study of aligning Covasim model outputs to observed data, operating from the assumption that combinations of uncertain model parameters could explain observed empirical phenomena. However, for policy planning purposes, it is also useful to consider what it *means* to calibrate an agent-based model in general: where our uncertainty comes from, and how that propagates to downstream policy interventions. We address two competing paradigms: the mean-focused approach that derives posterior parameter distributions expected to be invariant to the random mixing in the population, and the trajectory-oriented optimization approach, which addresses which combinations of model parameters and random number streams are capable of explaining the data and may offer a wider range of candidate parameterizations. In our case study, we are able to find suitable model parameterizations under both paradigms that are also able to explore potential (likely very effective) policy interventions. Thus, we argue that both viewpoints should be used to tame stochastic dynamical systems and inform decision making in uncertain and high-stakes contexts, such as epidemic control.

**Acknowledgments and Funding**
DOG and RAH are supported by Washington University Pilot Funds grant GF0012370. Covasim was developed at the Institute for Disease Modeling, part of the Bill & Melinda Gates Foundation, with contributions from others ([covasim.org](https://covasim.org)). The authors acknowledge the Research Infrastructure Services (RIS) group at Washington University in St. Louis for providing computational resources and services needed to generate the research results delivered within this paper. URL: [https://ris.wustl.edu](https://ris.wustl.edu)

# References

Abril-Pla, Oriol, Virgile Andreani, Colin Carroll, Larry Dong, Christopher J. Fonnesbeck, Maxim Kochurov, Ravin Kumar, Junpeng Lao, Christian C. Luhmann, Osvaldo A. Martin, Michael Osthege, Ricardo Vieira, Thomas Wiecki, and Robert Zinkov. 2023. "PyMC: A Modern, and Comprehensive Probabilistic Programming Framework in Python." *PeerJ Computer Science* 9:e1516. doi: 10.7717/peerj-cs.1516.

Akiba, Takuya, Shotaro Sano, Toshihiko Yanase, Takeru Ohta, and Masanori Koyama. 2019. "Optuna: A Next-Generation Hyperparameter Optimization Framework." Pp. 2623–31 in *Proceedings of the 25th ACM SIGKDD International Conference on Knowledge Discovery & Data Mining, KDD '19*. New York, NY, USA: Association for Computing Machinery.

Andrianakis, Ioannis, Ian R. Vernon, Nicky McCreesh, Trevelyan J. McKinley, Jeremy E. Oakley, Rebecca N. Nsubuga, Michael Goldstein, and Richard G. White. 2015. "Bayesian History Matching of Complex Infectious Disease Models Using Emulation: A Tutorial and a Case Study on HIV in Uganda." *PLOS Computational Biology* 11(1):e1003968. doi: 10.1371/journal.pcbi.1003968.

Aylett-Bullock, Joseph, Carolina Cuesta-Lazaro, Arnau Quera-Bofarull, Miguel Icaza-Lizaola, Aidan Sedgewick, Henry Truong, Aoife Curran, Edward Elliott, Tristan Caulfield, Kevin Fong, Ian Vernon, Julian Williams, Richard Bower, and Frank Krauss. 2021. "J UNE : Open-Source Individual-Based Epidemiology Simulation." *Royal Society Open Science* 8(7):210506. doi: 10.1098/rsos.210506.

Baker, Evan, Pierre Barbillon, Arindam Fadikar, Robert B. Gramacy, Radu Herbei, David Higdon, Jiangeng Huang, Leah R. Johnson, Pulong Ma, Anirban Mondal, Bianica Pires, Jerome Sacks, and Vadim Sokolov. 2020. "Analyzing Stochastic Computer Models: A Review with Opportunities." doi: 10.48550/ARXIV.2002.01321.

Beaumont, Mark A., Jean-Marie Cornuet, Jean-Michel Marin, and Christian P. Robert. 2009. "Adaptive Approximate Bayesian Computation." *Biometrika* 96(4):983–90.

Bedson, Jamie, Laura A. Skrip, Danielle Pedi, Sharon Abramowitz, Simone Carter, Mohamed F. Jalloh, Sebastian Funk, Nina Gobat, Tamara Giles-Vernick, Gerardo Chowell, João Rangel de Almeida, Rania Elessawi, Samuel V. Scarpino, Ross A. Hammond, Sylvie Briand, Joshua M. Epstein, Laurent Hébert-Dufresne, and Benjamin M. Althouse. 2021. "A Review and Agenda for Integrated Disease Models Including Social and Behavioural Factors." *Nature Human Behaviour* 5(7):834–46. doi: 10.1038/s41562-021-01136-2.

Bergstra, James, Rémi Bardenet, Yoshua Bengio, and Balázs Kégl. 2011. "Algorithms for Hyper-Parameter Optimization." in *Advances in Neural Information Processing Systems*. Vol. 24. Curran Associates, Inc.

Binois, Mickaël, Robert B. Gramacy, and Mike Ludkovski. 2018. "Practical Heteroscedastic Gaussian Process Modeling for Large Simulation Experiments." *Journal of*

*Computational and Graphical Statistics* 27(4):808–21. doi: 10.1080/10618600.2018.1458625.

Burke, Donald S., Joshua M. Epstein, Derek A. T. Cummings, Jon I. Parker, Kenneth C. Cline, Ramesh M. Singa, and Shubha Chakravarty. 2006. "Individual-Based Computational Modeling of Smallpox Epidemic Control Strategies." *Academic Emergency Medicine* 13(11):1142–49. doi: 10.1197/j.aem.2006.07.017.

Fadikar, Arindam, Mickael Binois, Nicholson Collier, Abby Stevens, Kok Ben Toh, and Jonathan Ozik. 2023. "Trajectory-Oriented Optimization of Stochastic Epidemiological Models."

Ferguson, N., D. Laydon, G. Nedjati Gilani, N. Imai, K. Ainslie, M. Baguelin, S. Bhatia, A. Boonyasiri, ZULMA Cucunuba Perez, G. Cuomo-Dannenburg, A. Dighe, I. Dorigatti, H. Fu, K. Gaythorpe, W. Green, A. Hamlet, W. Hinsley, L. Okell, S. Van Elsland, H. Thompson, R. Verity, E. Volz, H. Wang, Y. Wang, P. Walker, P. Winskill, C. Whittaker, C. Donnelly, S. Riley, and A. Ghani. 2020. *Report 9: Impact of Non-Pharmaceutical Interventions (NPIs) to Reduce COVID19 Mortality and Healthcare Demand*. Imperial College London. doi: 10.25561/77482.

Ferguson, Neil M., Derek A. T. Cummings, Christophe Fraser, James C. Cajka, Philip C. Cooley, and Donald S. Burke. 2006. "Strategies for Mitigating an Influenza Pandemic." *Nature* 442(7101):448–52. doi: 10.1038/nature04795.

Garnett, Roman. 2023. *Bayesian Optimization*. Cambridge University Press.

Germann, Timothy C., Kai Kadau, Ira M. Longini, and Catherine A. Macken. 2006. "Mitigation Strategies for Pandemic Influenza in the United States." *Proceedings of the National Academy of Sciences* 103(15):5935. doi: 10.1073/pnas.0601266103.

Gramacy, Robert B. 2020. *Surrogates: Gaussian Process Modeling, Design, and Optimization for the Applied Sciences*. CRC Press.

Halloran, M. Elizabeth, Neil M. Ferguson, Stephen Eubank, Ira M. Longini, Derek A. T. Cummings, Bryan Lewis, Shufu Xu, Christophe Fraser, Anil Vullikanti, Timothy C. Germann, Diane Wagener, Richard Beckman, Kai Kadau, Chris Barrett, Catherine A. Macken, Donald S. Burke, and Philip Cooley. 2008. "Modeling Targeted Layered Containment of an Influenza Pandemic in the United States." *Proceedings of the National Academy of Sciences* 105(12):4639–44. doi: 10.1073/pnas.0706849105.

Hammond, Ross A. 2015. "Considerations and Best Practices in Agent-Based Modeling to Inform Policy." in *Assessing the Use of Agent-Based Models for Tobacco Regulation*. National Academies Press (US).

Hammond, Ross A., and Shari Barkin. 2024. "Making Evidence Go Further: Advancing Synergy between Agent-Based Modeling and Randomized Control Trials." *Proceedings of the National Academy of Sciences* 121(21):e2314993121. doi: 10.1073/pnas.2314993121.

Hammond, Ross, Joseph T. Ornstein, Rob Purcell, Matthew D. Haslam, and Matt Kasman. 2021. "Modeling Robustness of COVID-19 Containment Policies."

Harville, David A. 1997. *Matrix Algebra From a Statistician's Perspective*. New York, NY: Springer.

Hinch, Robert, William J. M. Probert, Anel Nurtay, Michelle Kendall, Chris Wymant, Matthew Hall, Katrina Lythgoe, Ana Bulas Cruz, Lele Zhao, Andrea Stewart, Luca Ferretti, Daniel Montero, James Warren, Nicole Mather, Matthew Abueg, Neo Wu, Olivier Legat, Katie Bentley, Thomas Mead, Kelvin Van-Vuuren, Dylan Feldner-Busztin, Tommaso Ristori, Anthony Finkelstein, David G. Bonsall, Lucie Abeler-Dörner, and Christophe Fraser. 2021. "OpenABM-Covid19—An Agent-Based Model for Non-Pharmaceutical Interventions against COVID-19 Including Contact Tracing." *PLOS Computational Biology* 17(7):e1009146. doi: 10.1371/journal.pcbi.1009146.

Holthuijzen, Maike F., Robert B. Gramacy, Cayelan C. Carey, Dave M. Higdon, and R. Quinn Thomas. 2024. "Synthesizing Data Products, Mathematical Models, and Observational Measurements for Lake Temperature Forecasting."

Iskauskas, Andrew, Jamie A. Cohen, Danny Scarponi, Ian Vernon, Michael Goldstein, Daniel Klein, Richard G. White, and Nicky McCreesh. 2024. "Investigating Complex HPV Dynamics Using Emulation and History Matching."

Iskauskas, Andrew, Ian Vernon, Michael Goldstein, Danny Scarponi, Nicky McCreesh, Trevelyan J. McKinley, and Richard G. White. 2024. "Emulation and History Matching Using the **Hmer** Package." *Journal of Statistical Software* 109(10). doi: 10.18637/jss.v109.i10.

Kerr, Cliff C., Dina Mistry, Robyn M. Stuart, Katherine Rosenfeld, Gregory R. Hart, Rafael C. Núñez, Jamie A. Cohen, Prashanth Selvaraj, Romesh G. Abeysuriya, Michał Jastrzębski, Lauren George, Brittany Hagedorn, Jasmina Panovska-Griffiths, Meaghan Fagalde, Jeffrey Duchin, Michael Famulare, and Daniel J. Klein. 2021. "Controlling COVID-19 via Test-Trace-Quarantine." *Nature Communications* 12(1):2993. doi: 10.1038/s41467-021-23276-9.

Kerr, Cliff C., Robyn M. Stuart, Dina Mistry, Romesh G. Abeysuriya, Katherine Rosenfeld, Gregory R. Hart, Rafael C. Núñez, Jamie A. Cohen, Prashanth Selvaraj, Brittany Hagedorn, Lauren George, Michał Jastrzębski, Amanda S. Izzo, Greer Fowler, Anna Palmer, Dominic Delport, Nick Scott, Sherrie L. Kelly, Caroline S. Bennette, Bradley G. Wagner, Stewart T. Chang, Assaf P. Oron, Edward A. Wenger, Jasmina Panovska-Griffiths, Michael Famulare, and Daniel J. Klein. 2021. "Covasim: An Agent-Based Model of COVID-19 Dynamics and Interventions." *PLOS Computational Biology* 17(7):e1009149. doi: 10.1371/journal.pcbi.1009149.

Kerr, Cliff, Robyn M. Stuart, Romesh G. Abeysuriya, Jamie A. Cohen, Paula Sanz-Leon, Alina Muellenmeister, and Daniel Klein. n.d. "Starsim: Agent-Based Disease Modeling." *Starsim: A Fast, Flexible Toolbox for Agent-Based Modeling of Health and Disease. In Preparation.* Retrieved December 14, 2024 (https://starsim.org/).

Longini Ira M., Nizam Azhar, Xu Shufu, Ungchusak Kumnuan, Hanshaoworakul Wanna, Cummings Derek A. T., and Halloran M. Elizabeth. 2005. "Containing Pandemic Influenza at the Source." *Science* 309(5737):1083–87. doi: 10.1126/science.1115717.

Mistry, Dina, and C. C. Kerr. 2021. "Synthpops: Synthetic Contact Network Generation."

O'Gara, David. 2024a. "Archive of O'Gara et. al: 'Improving Policy-Oriented Agent-Based Modeling with History Matching: A Case Study.'"

O'Gara, David. 2024b. "Davidogara/hetGPy."

O'Gara, David, Samuel F. Rosenblatt, Laurent Hébert-Dufresne, Rob Purcell, Matt Kasman, and Ross A. Hammond. 2023. "TRACE-Omicron: Policy Counterfactuals to Inform Mitigation of COVID-19 Spread in the United States." *Advanced Theory and Simulations* 2300147. doi: 10.1002/adts.202300147.

Ozik, Jonathan, Justin M. Wozniak, Nicholson Collier, Charles M. Macal, and Mickaël Binois. 2021. "A Population Data-Driven Workflow for COVID-19 Modeling and Learning." *The International Journal of High Performance Computing Applications* 35(5):483–99. doi: 10.1177/10943420211035164.

Panovska-Griffiths, Jasmina, Thomas Bayley, Tony Ward, Akashaditya Das, Luca Imeneo, Cliff Kerr, and Simon Maskell. 2023. *Machine Learning Assisted Calibration of Stochastic Agent-Based Models for Pandemic Outbreak Analysis*. preprint. In Review. doi: 10.21203/rs.3.rs-2773605/v1.

Pukelsheim, Friedrich. 1994. "The Three Sigma Rule." *The American Statistician* 48(2):88–91. doi: 10.1080/00031305.1994.10476030.

Reiker, Theresa, Monica Golumbeanu, Andrew Shattock, Lydia Burgert, Thomas A. Smith, Sarah Filippi, Ewan Cameron, and Melissa A. Penny. 2021. "Emulator-Based Bayesian Optimization for Efficient Multi-Objective Calibration of an Individual-Based Model of Malaria." *Nature Communications* 12(1):7212. doi: 10.1038/s41467-021-27486-z.

Santner, Thomas J., Brian J. Williams, and William I. Notz. 2018. *The Design and Analysis of Computer Experiments*. New York, NY: Springer New York.

Shattock, Andrew J., Epke A. Le Rutte, Robert P. Dünner, Swapnoleena Sen, Sherrie L. Kelly, Nakul Chitnis, and Melissa A. Penny. 2022. "Impact of Vaccination and Non-Pharmaceutical Interventions on SARS-CoV-2 Dynamics in Switzerland." *Epidemics* 38:100535. doi: 10.1016/j.epidem.2021.100535.

Sisson, S. A., Y. Fan, and Mark M. Tanaka. 2007. "Sequential Monte Carlo without Likelihoods." *Proceedings of the National Academy of Sciences* 104(6):1760–65. doi: 10.1073/pnas.0607208104.

Sun, Furong, and Robert B. Gramacy. 2024. "Maximin: Space-Filling Design under Maximin Distance."

Vernon, I., J. Owen, J. Aylett-Bullock, C. Cuesta-Lazaro, J. Frawley, A. Quera-Bofarull, A. Sedgewick, D. Shi, H. Truong, M. Turner, J. Walker, T. Caulfield, K. Fong, and F. Krauss. 2022. "Bayesian Emulation and History Matching of JUNE." *Philosophical Transactions of the Royal Society A: Mathematical, Physical and Engineering Sciences* 380(2233):20220039. doi: 10.1098/rsta.2022.0039.

Vernon, Ian, Michael Goldstein, and Richard G. Bower. 2010. "Galaxy Formation: A Bayesian Uncertainty Analysis." *Bayesian Analysis* 5(4):619–69. doi: 10.1214/10-BA524.

Vernon, Ian, Junli Liu, Michael Goldstein, James Rowe, Jen Topping, and Keith Lindsey. 2018. "Bayesian Uncertainty Analysis for Complex Systems Biology Models: Emulation, Global Parameter Searches and Evaluation of Gene Functions." *BMC Systems Biology* 12(1):1. doi: 10.1186/s12918-017-0484-3.

Improving Policy-Oriented Agent-Based Modeling with History Matching: A Case Study
Supplementary Figures
David O'Gara et. al

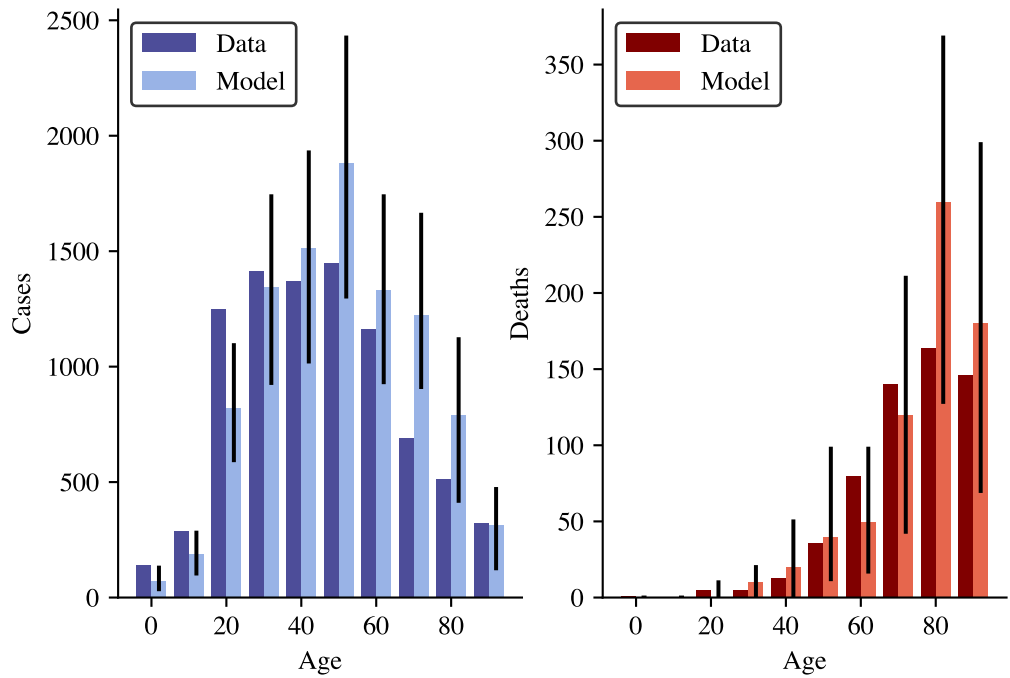

**Figure S1:** Histograms of diagnoses and deaths binned by age for observed data and calibrated Covasim outputs. Error bars represent 95% quantiles of outcomes across model runs.

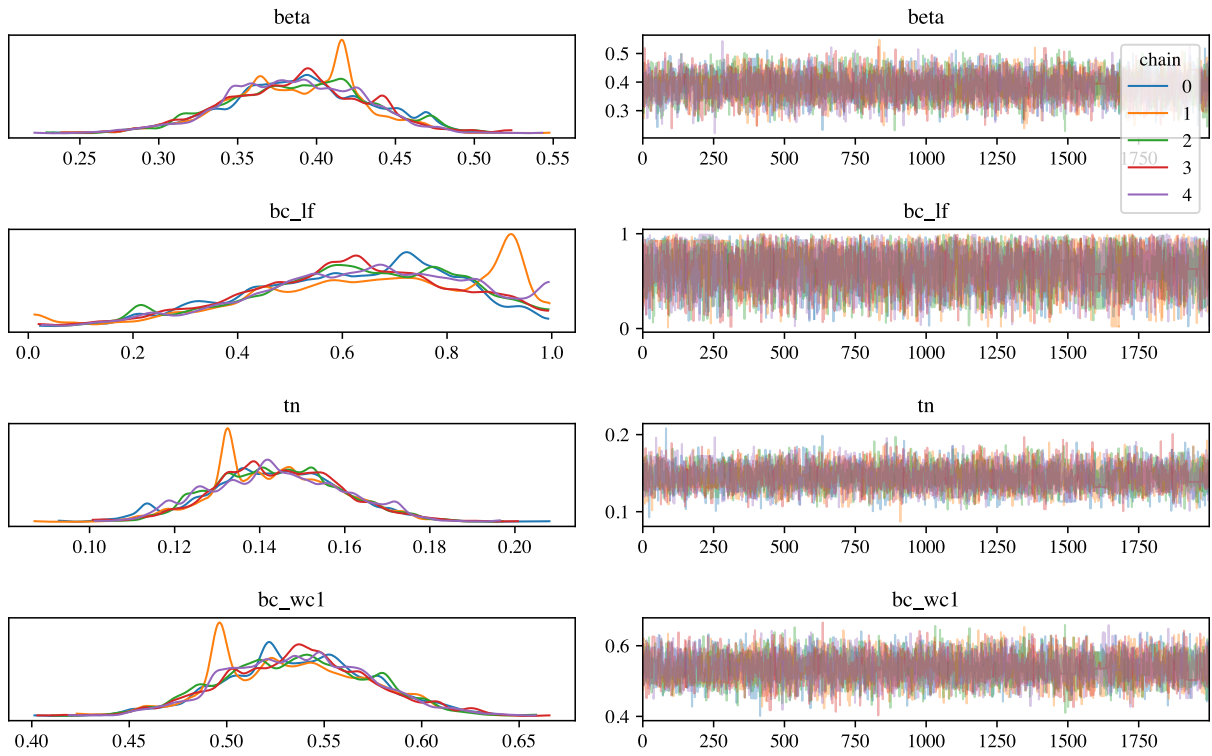

**Figure S2:** Markov chain Monte Carlo diagnostics for the ABC-SMC calibration for the time-series of diagnoses. Lefthand panels represent estimations of the posterior density for Covasim parameter values (scaled to be between 0 and 1). Righthand panels show the 5 chains for 2,000 MCMC samples.

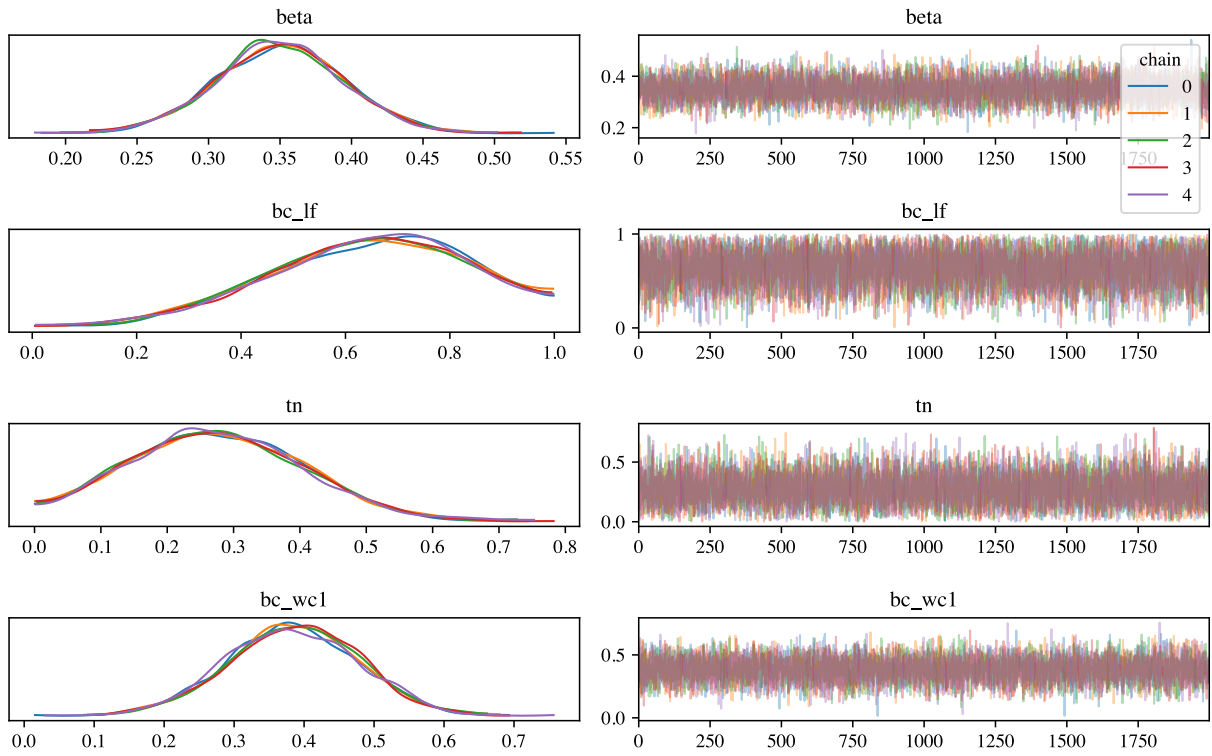

**Figure S3:** Markov chain Monte Carlo diagnostics for the ABC-SMC calibration for the time-series of deaths. Lefthand panels represent estimations of the posterior density for Covasim parameter values (scaled to be between 0 and 1). Righthand panels show the 5 chains for 2,000 MCMC samples.